\documentclass[runningheads]{llncs}
\usepackage[table]{xcolor}
\usepackage{graphicx}
\usepackage{multirow}
\usepackage{makecell}
\begin{document}
\title{Left Ventricle Contouring of Apical Three-Chamber Views on 2D Echocardiography\thanks{Supported by Ultromics Ltd.}}
\titlerunning{LV Contouring of 3-Chamber Echo Views}
%
\author{Alberto Gomez\inst{1, 2} \and
Mihaela Porumb\inst{1} \and
Angela Mumith\inst{1} \and
Thierry Judge\inst{3} \and
Shan Gao\inst{1} \and
Woo-Jin Cho Kim\inst{1} \and
Jorge Oliveira\inst{1} \and
Agis Chartsias\inst{1}}
%
\authorrunning{A. Gomez et al.}
%
\institute{Ultromics Ltd, Oxford, UK  \hspace{1cm}
King's College London, UK \and
Sherbrooke University, Canada \and
\email{alberto.gomez@ultromics.com}}
%
\maketitle              
\begin{abstract}
We propose a new method to automatically contour the left ventricle on 2D echocardiographic images. Unlike most existing segmentation methods, which are based on predicting segmentation masks, we focus at predicting the endocardial contour and the key landmark points within this contour (basal points and apex). This provides a representation that is closer to how experts perform manual annotations and hence produce results that are physiologically more plausible.

Our proposed method uses a two-headed network based on the U-Net architecture. One head predicts the 7 contour points, and the other head predicts a distance map to the contour. This approach was compared to the U-Net and to a point based approach, achieving performance gains of up to 30\% in terms of landmark localisation ($<4.5mm$) and distance to the ground truth contour ($<3.5mm$).

\end{abstract}
\section{Introduction}

Left ventricular (LV) contouring in echocardiography (echo) is a crucial step towards deriving cardiac measures including strain, LV volume, ejection fraction, cardiac output, and others. This task is typically carried out by manual delineation of B-mode images by experts. Such manual work is time consuming and prone to high inter- and intra- operator variability \cite{leclerc2019camus}. As a result, automation of LV contouring has been investigated thoroughly over the past decades. Since most of the aforementioned measures can be derived from contours obtained in two-chamber and four-chamber views, the three-chamber view (which is crucial for strain measures and accurate LV volumes) has been mostly neglected. In this paper, we focus our evaluation on contouring of the three-chamber view.

Most related work has aimed at solving the \emph{LV segmentation} problem by computing the closed region of the input image that includes the LV blood pool. The best performing methods to date use convolutional neural networks (CNNs) to predict the binary mask representing the ventricular blood pool from a frame or for every frame in a sequence. 

Leclerc et al. \cite{leclerc2019camus} proposed multiple encoder-decoder architectures to predict LV segmentation masks in the CAMUS dataset, which includes two-chamber and four-chamber views (but not three-chamber), in a frame-by-frame fashion. They found that U-Net \cite{ronneberger2015u} performed best for this task. Ouyang et al. \cite{ouyang2020echonet} proposed a two step process towards estimating the ejection fraction (EF). First, they trained a segmentation model on end-diastole (ED) and end-systole (ES) frames, and applied it to all frames in the cycle. Then, they combined the frame-wise predicted segmentation with the original sequence to regress EF using a spatio-temporal convolution model in the clip. Their spatio-temporal model requires resampling each cycle to a fixed length of 32 frames, losing potentially useful information about cardiac dynamics. An alternative way of exploiting the temporal domain is with the use of recurrent blocks. For example, Li et al. \cite{li2019recurrent} proposed an encoder-decoder architecture were the decoder was made of convolutional LSTM blocks, passing frame information to the next frame, but without any skip connections between the encoder and the decoder. Before them, Poudel et al.~\cite{Poudel2016} incorporated a recurrent block into the bottleneck of a U-Net, to leverage spatial consistency across slices in a cardiac magnetic resonance stack. In this paper, we incorporate a recurrent block into a U-Net inspired architecture but to leverage temporal consistency across consecutive frames. Recently, Painchaud et al. \cite{painchaud2022echocardiography} proposed a post-processing method to take frame-wise segmentations and correct temporal inconsistencies. 

All previous methods aimed at improving the quantification of area-based (or volume based) quantities, such as EF. This makes focusing on the prediction of closed segmentation masks appropriate. However, in many other applications, such as regional strain, cardiac motion, and shape analysis, predicting a endocardial contour open at the mitral valve plane would be a more sensible choice. Moreover, retrieving the location of anatomical landmarks along the contour, such as the basal points and the apex, are often required to split the ventricle into segments and to track specific anatomy throughout the cardiac cycle. Previous efforts aiming at directly predicting anatomical points along the contour are limited to the work by Porumb et al. \cite{porumb2021site}, where points along the contour of a two-chamber or a four-chamber view were regressed using a modified U-Net architecture using differentiable spatial to numerical transform (DSNT) layers \cite{nibali2018numerical}. The major drawbacks of this approach are that it assumes that all the landmarks can be consistently annotated by clinicians, when in fact only the 2 basal points and the apex are anatomically meaningful and can  be identified consistently, which is insufficient to accurately define a contour; and that it uses one frame at a time, being unable to leverage temporal consistency of the sequence.


In this paper, we propose a contouring method for 2D cardiac ultrasound sequences. The novel contributions are: 1) design of U-Net based multi-decoder for predicting the endocardial LV contour (instead of a segmentation mask) and three consistent anatomical landmarks in it: the basal points and the apex, 2) exploit frame interdependence by using a recurrent module and temporal regularization, and 3) propose contour evaluation metrics based on distances (rather than overlap measures) that are intuitive and clinically relevant.

\section{Methods}

\begin{figure}[!htb]
\includegraphics[width=\textwidth]{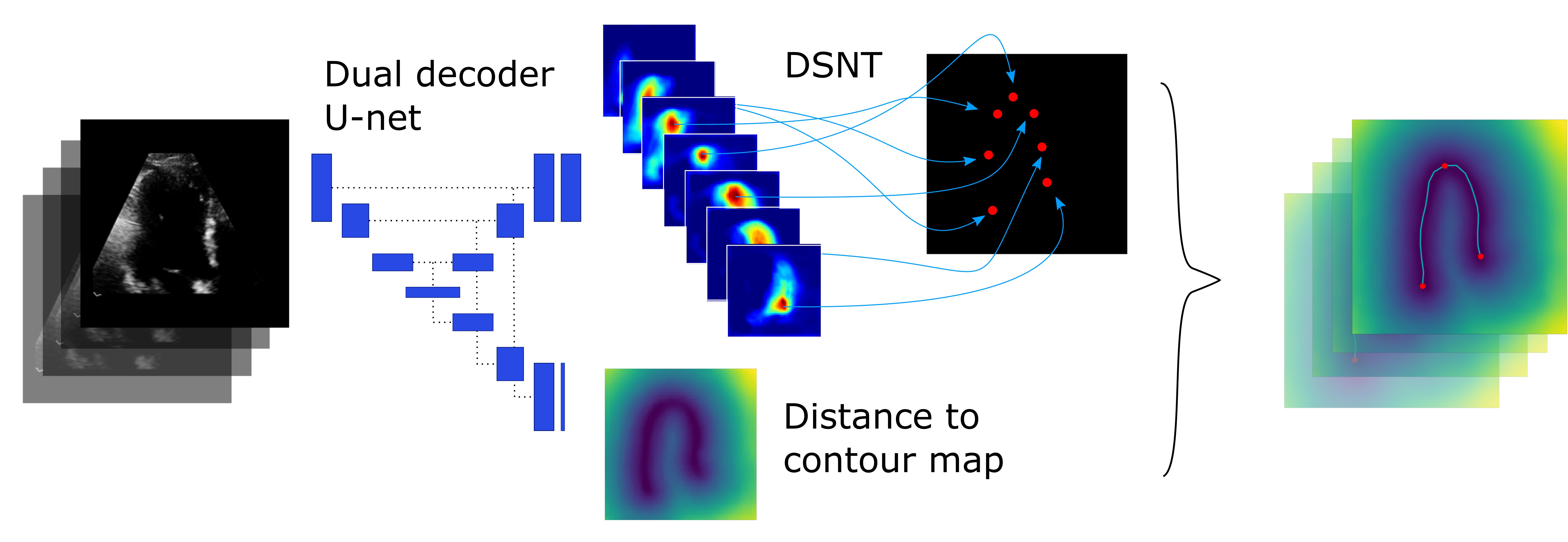}
\caption{Overview of the method. A dual-decoder U-Net processes the input sequence predicting, for each frame, the location of 7 contour points (as heatmaps, on the top decoder) and a distance-to-contour map (on the bottom decoder). Using DSNT layers~\cite{nibali2018numerical}, point coordinates are calculated for each heatmap. The contour is calculated as the spline passing by all 7 predicted points through the shortest geodesic distance on the distance map.} \label{fig:overview}
\end{figure}

We designed a neural network that takes as input a fixed-length sequence of three-chamber ultrasound images. This network predicts  a distance-to-contour map and the location of $N_p$ points on the contour for each frame. The contour curve is derived as the shortest geodesic path passing by the predicted points, using the predicted distance map to define geodesic distance. An overview of the method is illustrated in Fig. \ref{fig:overview}, and each step is described in turn in this section.

\subsection{Data preparation} 

%

Data belongs to an institutional database and include 525 scans from 460 patients that are part of EVAREST~\cite{woodward2022real}, a UK multi-site, multi-vendor clinical trial, 
obtained using GE, Phillips or Siemens ultrasound systems. The dataset contains echocardiograms from patients diagnosed with coronary artery disease, COVID patients or healthy volunteers. The EVAREST trial specifically, contains echocardiograms obtained from patients at rest or from patients subjected to exercise or pharmacological stress. Heart rates vary between 40 and 280 and the number of frames per scan is between 6 and 130. For each patient there was at least one 3 chamber sequence stored as DICOM, containing one or more cardiac cycles. The patients were split into $T=440$ for training and 85 for validation. The data were anonymised and pre-processed so that most text and annotation information outside the ultrasound region is removed. 

The data was annotated by expert physiologists using in-house software, as follows. For every clip, the operator identified the end-diastole (ED) and end-systole (ES) frames in a cycle of their choice. Both frames were in turn annotated by selecting 7 points, of which 3 corresponded to the two basal points (BPs) and the apex (AP). The other 4 points were picked, two on each side of the AP, along the contour and roughly equidistant between the AP and the corresponding BP, as exemplified in Fig. \ref{fig:overview}. Operators can add additional points if needed. As a result, only the AP and BPs correspond to anatomical landmarks and the other points serve to define the contour. The ground truth contour is computed as a smooth spline that passes by the selected points.

We converted the ground truth contour information into distance maps by defining a pixel grid, of the same size as the input images, and computing at each pixel the shortest distance, in pixels, to the contour (e.g. Fig. \ref{fig:overview}). The ground truth labels $L_i=(\{P\}_j, D)_i$ for each image $i$ are tuples including 7 points $\{P\}_j$ (of which the first and last are BPs and the mid point is the apex), and the distance-to-contour map $D$. $D$ is normalised by the half-diagonal of the image.

During training, a sequence of length 15 frames was extracted from each clip. The sequence was extracted at random (sampling uniformly) at each epoch, while ensuring that at least 1 frame within was annotated. The data component of the loss (c.f. Sec \ref{sec:method1}) was computed only on the annotated frames. 

\subsection{Prediction of the distance-to-contour map and contour points}
\label{sec:method1}

\begin{figure}[!htb]
\includegraphics[width=\textwidth]{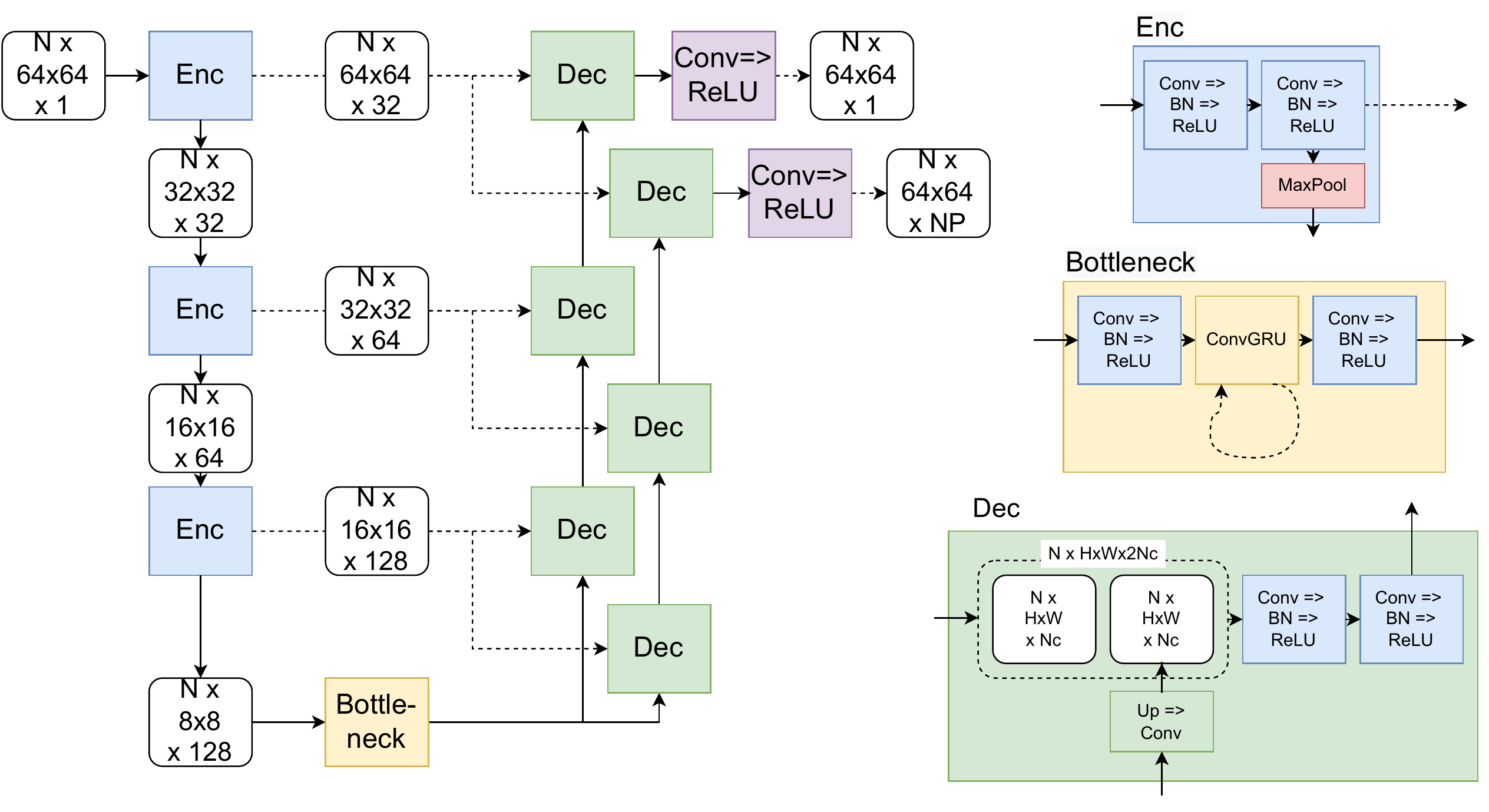}
\caption{Architecture of the contour prediction network, on the left, and detail of the encoder (Enc), bottleneck and decoder (Dec) blocks, on the right. The architecture is based on a U-Net but with two parallel decoder branches, the first one used to predict the distance-to-contour map and the second one (with $NP=7$ output channels) to predict the location of $NP=7$ contour points. On the right, detail of the model blocks.} \label{fig:cnn}
\end{figure}

We propose a dual encoder U-Net based architecture to jointly predict contour points $\{\tilde{P}\}_j$ and distance-to-contour maps $\tilde{D}$.  An overview of the neural network architecture is illustrated in Fig. \ref{fig:cnn}.

The input to the network is an image clip of $N=15$ frames and size of $64\times64$ pixels. The output of the network is a tensor of size $15 \times 64 \times 64 \times 8$, where the 8 output channels correspond to one distance-to-contour map and 7 heat-maps predicting the location of the 7 points. These are named, respectively $\tilde{D}$ and $\{\tilde{H}\}_{j=0,...,6}$. The predicted location of the points can be computed using a Differentiable Spatial to Numerical Transform \cite{nibali2018numerical} layer (DSNT). This layer computes point coordinates from a heat-map as a spatial average weighted by normalised coordinates, as $\tilde{P}_j = DSTN(\tilde{H}_j)$. The loss function to train the model was composed of a composite data fit term, $L_D$, and a regularization term, $L_R$. The data term for $N_p=7$ points is:
\begin{equation}
\begin{array}{c}
    L_D = L_m + L_p \textnormal{, where} \\
    L_m = \frac{1}{T} \sum_i MAE(D_i, \tilde{D}_i) \\
    L_p = \frac{1}{T} \sum_i \frac{1}{N_p}\sum_j ||P_{j,i}, \tilde{P}_{j,i}||^2.
    \end{array}
\end{equation}
The regularization term penalises the temporal gradient of the predicted sequence as follows:
\begin{equation}
    L_R = \frac{1}{T} \sum_i \left(|\nabla_t \tilde{D}_i| + \frac{1}{N_p}\sum_j |\nabla_t  \tilde{H}_{j,i}| \right),
\end{equation}
where the temporal gradient operator is approximated by first order time difference. The final loss is $L=L_D + \lambda L_R$, where the optimal value of $\lambda=0.001$ was found by hyper-parameter search in the interval $[0, 10^{-4}, 10^{-3}, 10^{-2}, 10^{-1}, 0.5, 1]$.

\subsection{Contour curve from distance-to-contour and points}

The contour is computed as a parametric smooth spline curve from the predicted points and the predicted distance-to-contour map by finding the shortest geodesic path from predicted basal point to basal point, passing by all predicted points, and using the distance map as a cost for computing the geodesic distance. 


We solve this by using the Dijkstra algorithm \cite{dijkstra1959note} between every pair of consecutive points in the contour. Finally, we sample the resulting parametric curve to 21 points (the 2 BPs, the AP, and 9 points uniformly sampled at each side of the AP).

\section{Implementation and Experiments}

\subsection{Implementation details}

The method was implemented in Python 3, using Tensorflow 2.7 for the deep learning pipeline. Training was carried out using a NVidia GeForce RTX 2080 Ti GPU. All models were trained using the Adam optimizer with a learning rate of $10^{-4}$ and a batch size of 4 for 150 epochs, which was in all cases well beyond convergence. The model with the lowest validation loss was used for evaluation.

\subsection{Experiments}

The aim of our experiments was to assess the accuracy of the predicted landmark points (using the Euclidean distance between predicted and ground truth points), and  the accuracy of the predicted contour as a whole (as the distance between 18 points sampled between each basal point and the apex in the predicted contour, to the ground truth contour). We report average and standard deviation values using the point-based method in \cite{porumb2021site} as a baseline, and the proposed method without (Proposed) and with (Proposed GRU) a Gated Recurrent Unit -GRU module in the bottleneck. 

Additionally, we provided experts with pairs of animations showing the contours over the cardiac cycle, and asked them to pick the most temporally consistent, while being blinded to the method used for each animation (which was randomized between baseline, proposed method without recurrent module, and whole proposed method). We then reported the frequency at which each method was picked best of the two.

\section{Results}
The results on accuracy of predicted landmarks and contours are shown in Table \ref{tab1}. The proposed method provides best performance both in accuracy of predicted landmarks  (up to a 30\% better than the baseline) and in terms of distance to contour (up to 4.7\% better than the baseline). Largest performance gains are in the landmark points, and particularly at the basal points. Using a recurrent bottleneck does match or improve the results further in all cases. 

\begin{table}[!htb]
\centering
\caption{Performance summary using two metrics: Point distance, average $\pm$ standard deviation of the Euclidean distance reported for the landmark points (basal point 1 and 2 -BP1 and BP2- and apex), and average values over those three landmarks; and  distance to contour, distance from 21 uniformly sampled points in the predicted contour to the ground truth contour. The relative improvement in terms of average accuracy is presented in green.}\label{tab1}
\begin{tabular}{ll|c|c|c}
                                &        & PointReg        & Proposed & Proposed GRU       \\ \hline \hline
\multirow{4}{*}{\rotatebox[origin=c]{90}{\makecell{
           Point \\ distance}}} & BP1   &  $5.44 \pm 1.50$ & $ 4.20\pm1.34 $ \color{green!60!black}{+22.8\%} &  $4.22\pm1.34 $ \color{green!60!black}{+22.5\%}\\
                                & Apex  &  $5.66 \pm 1.20$ & $ 4.88\pm1.36 $ \color{green!60!black}{+13.8\%} &  $ 4.80\pm1.14 $ \color{green!60!black}{+15.2\%}\\
                                & BP2   &  $6.36 \pm 1.90$ & $ 5.29\pm1.38 \color{green!60!black}{+16.8\%}$ &  $4.45\pm1.21 $ \color{green!60!black}{+30.0\%}\\ 
                     & \textit{Average} &  $5.82 \pm 1.53$  &  $ 4.79\pm 1.36$ \color{green!60!black}{+17.7\%}&  $ 4.49\pm1.23 $ \color{green!60!black}{+22.9\%}\\ \hline \hline
\multicolumn{2}{l|}{\makecell{Distance \\ to contour}}  &   $ 3.64 \pm 1.49$ &  $ 3.57\pm1.45 $ \color{green!60!black}{+1.9\%}&  $3.47 \pm1.28 $ \color{green!60!black}{+4.7\%}\\ \hline
\end{tabular}
\end{table}

\begin{figure}[!htb]
\centering
\includegraphics[width=.8\textwidth]{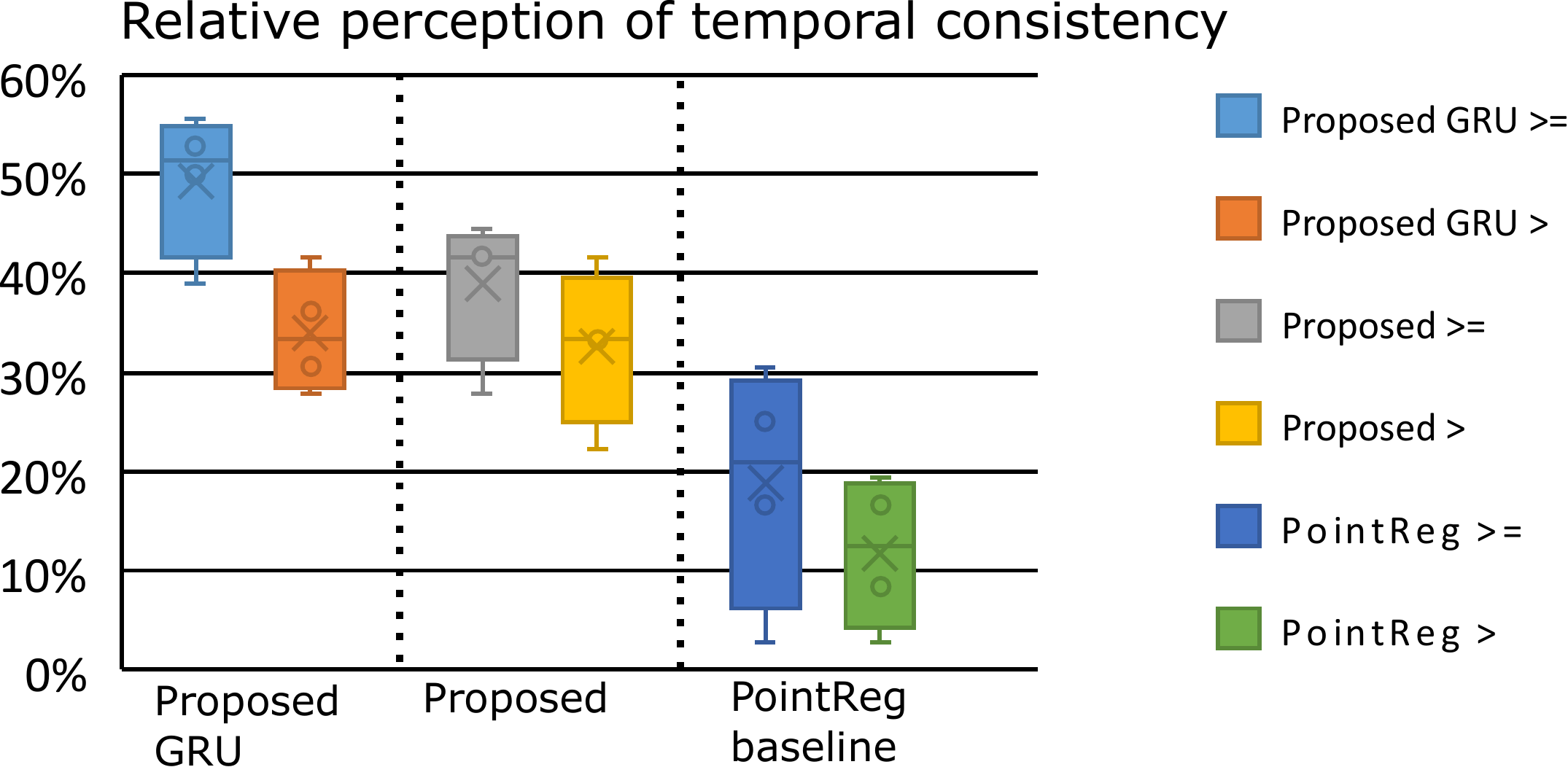}
\caption{Relative temporal consistency of the three competing methods. From left to right, cases where experts judged: 1) the proposed method equal or better; 2) or just  better; 3) the proposed method without the GRU bottleneck equal or better; 4) or just better; 5) the baseline (PointReg) equal or better; 6) or just better. In all cases, comparisons are performed against the other two methods.} \label{fig:tempcons}
\end{figure}

The qualitative results on temporal consistency are reported in Fig \ref{fig:tempcons}. The box and whiskers plots show the compiled scores from 5 experts, indicating the fraction of the assessed cases where each method was better or equal, in terms of temporal consistency, than its alternative (indicated $>=$ in the figure), or the fraction where it was better than the alternative (indicated $>$). The proposed method, when using a recurrent bottleneck, performs better or equal than any of the other two methods in over 50\% of the cases, compared to the baseline method which was better or equal to the alternative method in just over 20\% of the cases.

\begin{figure}[!htb]
\includegraphics[width=\textwidth]{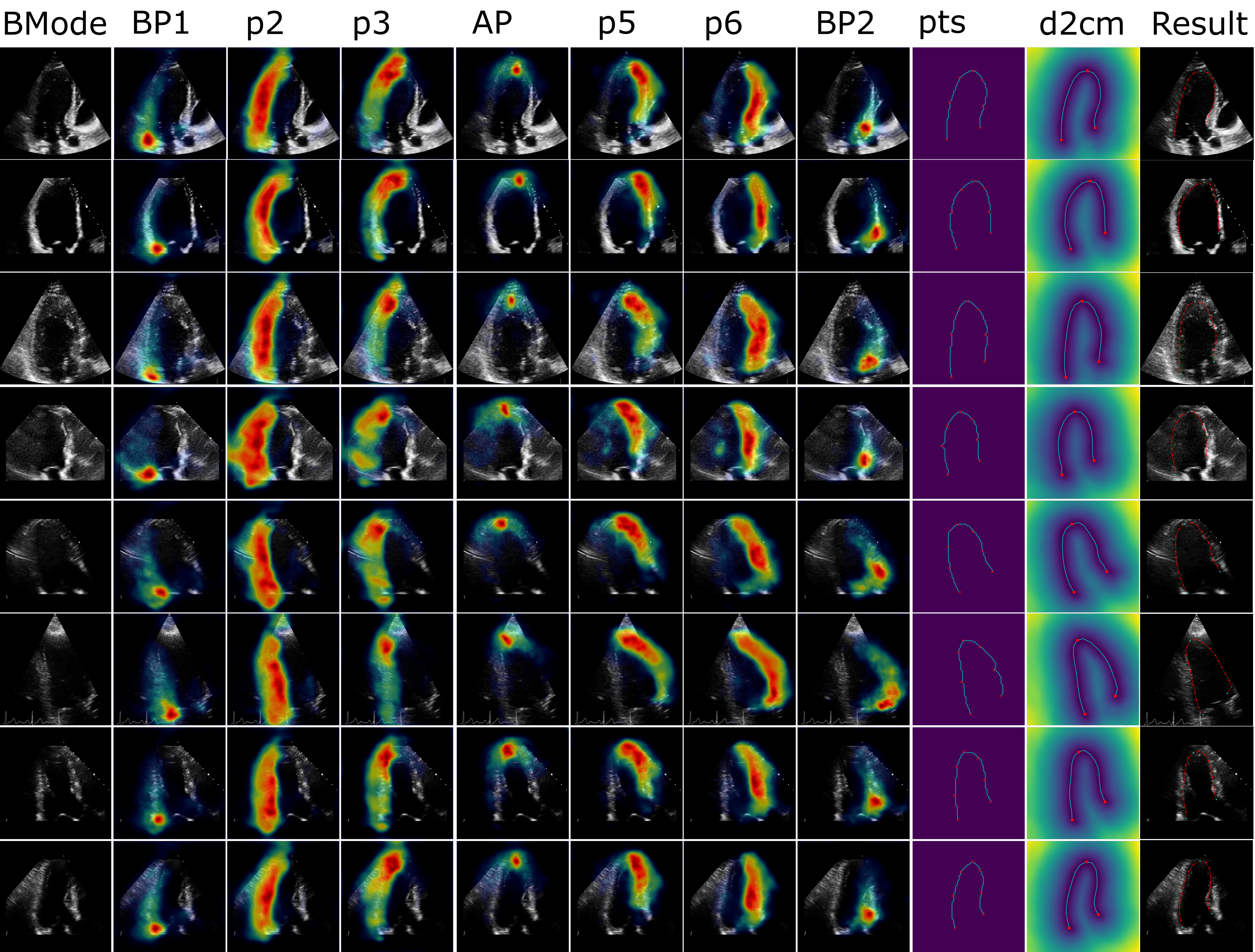}
\caption{Example intermediate and final outputs for a frame from eight subjects of different quality representative of the validation set. Over the columns: the original B-mode frame, the point location heat-maps, and the predicted contour (shortest geodesic path, fitted spline and result). Refer to the text for further detail.} \label{fig:example_out}
\end{figure}

Last, we present 8 contour examples, using the our model with a GRU block, in Fig. \ref{fig:example_out}. Each of the 8 rows represents a frame from 8 different subjects. The first column shows the input frame; Columns 2 to 8 show the predicted heatmaps corresponding to the 7 points, where it can be seen that the three landmarks (cols. 2 and 8: BPs and col 5: AP) are consistently predicted with much greater accuracy than the non landmark points (p2, p3, p5, p6). The last 3 columns show, respectively, the shortest geodesic path passing by the 7 points (pts), the smooth spline passing by the landmarks on the predicted distance to contour map (d2cm), and the resulting contour overlaid onto the original frame (result).

\section{Discussion and Conclusions}

We have presented a method to predict endocardial LV contours from B-mode sequences acquired on the 3-chamber view. The proposed method computes the contour from three predicted landmarks and a predicted distance-to-contour map, as opposed to related work where a binary mask of the LV is predicted. The contour based approach has the advantage of consistent retrieval of anatomical landmarks (the basal points and the apex) that are normally needed to compute clinically relevant metrics that involve splitting the LV into segments, or having an accurate delimitation of the LV at the mitral valve plane. 

The proposed method leverages temporal information using a GRU module at the bottleneck, which improves landmark detection and provides  more time-consistent contours. Specifically, our method outperforms the baseline point regression method in \cite{porumb2021site}, even without the GRU. Results shown in Fig. \ref{fig:example_out} suggest that a reason for this is the lack of consistency in annotations of the non-landmark points, which in turn cannot be predicted accurately, and the model seems unsure of where to locate these intermediate points along the contour.

In this paper, we have also proposed to move from shape overlap metrics, such as Dice and intersection-over-union, to more intuitive geometric measures, specifically Euclidean distance to landmark points and average contour to contour distance. 
The results reported by these metrics seem to correlate well with the qualitative assessment of the resulting contours over time.

%
%
%

Our model achieves an average accuracy of up to 4.49mm for the landmarks and 3.47 mm for the contour itself. Leclerc et al. \cite{leclerc2019camus} found that inter-user variability was up to $4.0 \pm 2.0$ mm (restricting the analysis to images of good and medium quality) which suggests that our results are of clinical use quality. Future work will involve more experiments with clinical users to  confirm this.

%
%
%
\bibliographystyle{splncs04}
\bibliography{bibliography}
%




\end{document}